\documentclass[12pt,draftclsnofoot,onecolumn]{IEEEtran}
\usepackage{graphicx}
\usepackage{latexsym}
\usepackage{amsmath, amstext,amssymb, amsfonts,amsbsy}
\usepackage{epsf}
\newtheorem{thm}{Theorem}[section]

\newtheorem{lem}[thm]{Lemma}
\newtheorem{rem}[thm]{Remark}

\newcommand{\figlbl}[1]{\label{fig:#1}}
\newcommand{\figref}[1]{Fig.~\ref{fig:#1}}

\newcounter{rmnum}
\newenvironment{romannum}{\begin{list}{{\upshape (\roman{rmnum})}}{\usecounter{rmnum}
\setlength{\leftmargin}{10pt} \setlength{\rightmargin}{8pt}
\setlength{\itemindent}{-1pt} }}{\end{list}}

\begin{document}
\title{Algorithms for Dynamic Spectrum Access with Learning for Cognitive Radio}
\author{{Jayakrishnan Unnikrishnan,~\IEEEmembership{Student Member,~IEEE,}
        and~Venugopal~V.~Veeravalli,~\IEEEmembership{Fellow,~IEEE}} \thanks{
        Copyright (c) 2008 IEEE. Personal use of this material is permitted. However, permission to use this material for any other purposes must be obtained from the IEEE by sending a request to pubs-permissions@ieee.org.
        This work was supported by a Vodafone Foundation Graduate Fellowship and by NSF Grant
CCF 07-29031, through the University of Illinois. This paper
was presented
        in part at the 47$^\text{th}$ IEEE Conference on Decision and Control, Cancun, Mexico, 2008, and at
        the Asilomar Conference on Signals, Systems and Computers, Pacific Grove, CA,
        2008.

The authors are with the Coordinated Science Laboratory and
Dept. of Electrical and Computer Engineering, University of
Illinois at Urbana- Champaign, Urbana, IL 61801, USA. Email
\{junnikr2, vvv\}@illinois.edu} }

\maketitle

\begin{abstract}
We study the problem of dynamic spectrum sensing and access in
cognitive radio systems as a partially observed Markov decision
process (POMDP). A group of cognitive users cooperatively tries
to exploit vacancies in primary (licensed) channels whose
occupancies follow a Markovian evolution. We first consider the
scenario where the cognitive users have perfect knowledge of
the distribution of the signals they receive from the primary
users. For this problem, we obtain a greedy channel selection
and access policy that maximizes the instantaneous reward,
while satisfying a constraint on the probability of interfering
with licensed transmissions. We also derive an analytical
universal upper bound on the performance of the optimal policy.
Through simulation, we show that our scheme achieves good
performance relative to the upper bound and improved
performance relative to an existing scheme.


We then consider the more practical scenario where the exact
distribution of the signal from the primary is unknown. We
assume a parametric model for the distribution and develop an
algorithm that can learn the true distribution, still
guaranteeing the constraint on the interference probability. We
show that this algorithm outperforms the naive design that
assumes a worst case value for the parameter. We also provide a
proof for the convergence of the learning algorithm.
\end{abstract}


\section{Introduction}
Cognitive radios that exploit vacancies in the licensed
spectrum have been proposed as a solution to the
ever-increasing demand for radio spectrum. The idea is to sense
times when a specific licensed band is not used at a particular
place and use this band for unlicensed transmissions without
causing interference to the licensed user (referred to as the
`primary'). An important part of designing such systems is to
develop an efficient channel selection policy. The cognitive
radio (also called the `secondary user') needs to adopt the
best strategy for selecting channels for sensing and access.
The sensing and access policies should jointly ensure that the
probability of interfering with the primary's transmission
meets a given constraint.


In the first part of this paper, we consider the design of such
a joint sensing and access policy, assuming a Markovian model
for the primary spectrum usage on the channels being monitored.
The secondary users use the observations made in each slot to
track the probability of occupancy of the different channels.
We obtain a suboptimal solution to the resultant POMDP problem.



In the second part of the paper, we propose and study a more
practical problem that arises when the secondary users are not
aware of the exact distribution of the signals that they
receive from the primary transmitters.
We develop an algorithm that learns these unknown statistics and show that this
scheme gives improved performance over the naive scheme that assumes
a worst-case value for the unknown distribution.

\subsection{Contribution}
When the statistics of the signals from the primary are known,
we show that, under our formulation, the dynamic spectrum
access problem with a group of cooperating secondary users is
equivalent in structure to a single user problem. We also
obtain a new analytical upper bound on the expected reward
under the optimal scheme. Our suboptimal solution to the POMDP
is shown via simulations to yield a performance that is close
to the upper bound and better than
that under an existing scheme. 

The main contribution of this paper is the formulation and
solution of the problem studied in the second part involving
unknown observation statistics. We show that unknown statistics
of the primary signals can be learned and provide an algorithm
that learns these statistics online and maximizes the expected
reward still satisfying a constraint on interference
probability.

\subsection{Related Work}
In most of the existing schemes \cite{zhao_krish_liu},
\cite{javidi_krish_zhao_liu} in the literature on dynamic
spectrum access for cognitive radios, the authors assume that
every time a secondary user senses a primary channel, it can
determine whether or not the channel is occupied by the
primary. A different scheme was proposed in
\cite{zhao_tong_swami} and \cite{chen_zhao_swami} where the
authors assume that the secondary transmitter receives
error-free ACK signals from the secondary's receivers whenever
their transmission is successful. The secondary users use these
ACK signals to track the channel states of the primary
channels. We adopt a different strategy in this paper. We
assume that every time the secondary users sense a channel they
see a random observation whose distribution depends on the
state of the channel. Our approach is distinctly different from
and more realistic than that in \cite{zhao_krish_liu,
javidi_krish_zhao_liu} since we do not assume that the
secondary users know the primary channel states perfectly
through sensing. We provide a detailed comparison of our
approach with that of \cite{zhao_tong_swami} and
\cite{chen_zhao_swami} after presenting our solution. In
particular, we point out that while using the scheme of
\cite{chen_zhao_swami} there are some practical difficulties in
maintaining synchronization between the secondary transmitter
and receiver. Our scheme provides a way around this difficulty,
albeit we require a dedicated control channel between the
secondary transmitter and receiver.

The problem studied in the second part of this paper that
involves learning of unknown observation statistics is new.
However, the idea of combining learning and dynamic spectrum
access was also used in \cite{motamedi_bahai} where the authors
propose a reinforcement-learning scheme for learning channel
idling probabilities and interference probabilities.

We introduce the basic spectrum sensing and access problem in
Section \ref{probform} and describe our proposed solution in
Section \ref{dp}. In Section \ref{unknown}, we elaborate on the
problem where the distributions of the observations are
unknown.  We present simulation results and comparisons with
some existing schemes in Section \ref{sims}, and our
conclusions in Section \ref{comments}.

%
%
%
%
%

\section{Problem Statement} \label{probform}
We consider a slotted system where a group of secondary users
monitor a set $\mathcal{C}$ of primary channels. The state of
each primary channel switches between `occupied' and
`unoccupied' according to the evolution of a Markov chain. The
secondary users can cooperatively sense any one out of the
channels in $\mathcal{C}$ in each slot, and can access any one
of the $L = |\mathcal{C}|$ channels in the same slot. In each
slot, the secondary users must satisfy a strict constraint on
the probability of interfering with potential primary
transmissions on any channel. When the secondary users access a
channel that is free during a given time slot, they receive a
reward proportional to the bandwidth of the channel that they
access. The objective of the secondary users is to select the
channels for sensing and access in each slot in such a way that
their total expected reward accrued over all slots is maximized
subject to the constraint on interfering with potential primary
transmissions every time they access a channel\footnote{We do
not consider scheduling policies in this paper and assume that
the secondary users have some predetermined scheduling policy
to decide which user accesses the primary channel every time
they determine that a channel is free for access.}. Since the
secondary users do not have explicit knowledge of the states of
the channels, the resultant problem is a constrained partially
observable Markov decision process (POMDP) problem.

We assume that all channels in $\mathcal{C}$ have equal
bandwidth $B$, and are statistically identical and independent
in terms of primary usage. The occupancy of each channel
follows a stationary Markov chain. The state of channel $a$ in
any time slot $k$ is represented by variable $S_a(k)$ and could
be either $1$ or $0$, where state $0$ corresponds to the
channel being free for secondary access and $1$
corresponds to the channel being occupied by some primary user. 



The secondary system includes a decision center that has access
to all the observations made by the cooperating secondary
users\footnote{The scheme proposed in this paper and the
analyses presented in this paper are valid even if the
cooperating secondary users transmit quantized versions of
their observations to the fusion center. Minor changes are
required to account for the discrete nature of the
observations.}. The observations are transmitted to the
decision center over a dedicated control channel. The same
dedicated channel can also be used to maintain synchronization
between the secondary transmitter and secondary receiver so
that the receiver can tune to the correct channel to receive
transmissions from the transmitter. The sensing and access
decisions in each slot are made at this decision center. When
channel $a$ is sensed in slot $k$, we use $\underline{X}_a(k)$
to denote the vector of observations made by the different
cooperating users on channel $a$ in slot $k$. These
observations represent the sampled outputs of the wireless
receivers tuned to channel $a$ that are employed by the
cognitive users. The statistics of these observations are
assumed to be time-invariant and distinct for different channel
states. The observations on channel $a$ in slot $k$ have
distinct joint probability density functions $f_0$ and $f_1$
when $S_a(k) = 0$ and $S_a(k) = 1$ respectively. The collection
of all observations up to slot $k$ is denoted by
$\underline{X}^{k}$, and the collection of observations on
channel $a$ up to slot $k$ is denoted by $\underline{X}_a^{k}$.
The channel sensed in slot $k$ is denoted by $u_k$, the
sequence of channels sensed up to slot $k$ is denoted by $u^k$,
and the set of time slots up to slot $k$ when channel $a$ was
sensed is denoted by $K_a^k$. The decision to access channel
$a$ in slot $k$ is denoted by a binary variable $\delta_a(k)$,
which takes value $1$ when channel $a$ is accessed in slot $k$,
and $0$ otherwise.

Whenever the secondary users access a free channel in some time
slot $k$, they get a reward $B$ equal to the bandwidth of each
channel in $\mathcal{C}$. The secondary users should satisfy
the following constraint on the probability of interfering with
the primary transmissions in each slot:
\[
\mathsf{P}(\{\delta_a(k) = 1 \} | \{S_a(k) = 1\}) \leq \zeta.
\]
In order to simplify the structure of the access policy, we
also assume that in each slot the decision to access a channel
is made using only the observations made in that slot. Hence it
follows that in each slot $k$, the secondary users can access
only the channel they sense in slot $k$, say channel $a$.
Furthermore, the access decision must be based on a binary
hypothesis test \cite{poor} between the two possible states of
channel $a$, performed on the observation $\underline{X}_a(k)$.
This leads to an access policy with a structure similar to that
established in \cite{chen_zhao_swami}. The optimal test
\cite{poor} is to compare the joint log-likelihood ratio (LLR)
$\mathcal{L}(\underline{X}_a(k))$ given by,
\[\mathcal{L}(\underline{X}_a(k)) = \log \left(
\frac{f_1(\underline{X}_a(k))}{f_0(\underline{X}_a(k))}\right)\] to some threshold $\Delta$
that is chosen to satisfy,
\begin{equation}
\mathsf{P}\left(\left\{ \mathcal{L}(\underline{X}_a(k)) <
\Delta\right\} | \{S_a(k) = 1\}\right) = \zeta \label{Delta}
\end{equation}
and the optimal access decision would be to access the sensed channel whenever the threshold
exceeds the joint LLR. Hence
\begin{equation}
\delta_a(k) = \mathcal{I}_{\left\{ \mathcal{L}(\underline{X}_a(k)) <
\Delta\right\}} \mathcal{I}_{\{u_k = a \}} \label{access}
\end{equation}
and the reward obtained in slot $k$ can be expressed as,
\begin{eqnarray}
\hat{r}_k&=& B\mathcal{I}_{\left\{S_{u_k}(k) = 0
\right\}}\mathcal{I}_{\left\{\mathcal{L}(\underline{X}_{u_k}(k)) <
\Delta\right\}} \label{rewardactual}
\end{eqnarray}
where $\mathcal{I}_E$ represents the indicator function of event $E$. The main advantage of the
structure of the access policy given in (\ref{access}) is that we can obtain a simple
sufficient statistic for the resultant POMDP without having to keep track of all the past
observations, as discussed later. It also has the added advantage \cite{chen_zhao_swami} that
the secondary users can set the thresholds $\Delta$ to meet the constraint on the probability
of interfering with the primary transmissions without relying on their knowledge of the Markov
statistics. 

Our objective is to generate a policy that makes optimal use of primary spectrum subject to the
interference constraint. We introduce a discount factor $\alpha \in (0,1)$ and aim
to solve the infinite horizon dynamic program with discounted rewards \cite{bertsekas}. That is, we seek the
sequence of channels $\{u_0, u_1, \ldots \}$, such that the $\displaystyle \sum_{k=0}^{\infty}
\alpha^k \mathsf{E} [\hat{r}_k]$ is maximized, where the expectation is performed over the
random observations and channel state realizations. We can show the following relation based
on the assumption of identical channels:
\begin{eqnarray}
\mathsf{E} [\hat{r}_k] &=& \mathsf{E}
\left[B\mathcal{I}_{\left\{S_{u_k}(k) = 0
\right\}}\mathcal{I}_{\left\{\mathcal{L}(\underline{X}_{u_k}(k)) <
\Delta\right\}}\right]\nonumber\\
&=& \mathsf{E} \left[ \mathsf{E}
\left[B\mathcal{I}_{\left\{S_{u_k}(k) = 0
\right\}}\mathcal{I}_{\left\{\mathcal{L}(\underline{X}_{u_k}(k)) <
\Delta\right\}} | S_{u_k}(k)\right]\right] \nonumber\\
&=& \mathsf{E} \left[ B(1-\hat{\epsilon})\mathcal{I}_{\{S_{u_k}(k) = 0 \}} \right] \label{exprewardnew}
\end{eqnarray}
where,
\begin{equation}
\hat{\epsilon} = \mathsf{P}(\{\mathcal{L}(\underline{X}_a(k)) >
\Delta\}| \{S_a(k) =
0\}). \label{epsilonorig} 
\end{equation}
Since all the channels are assumed to be identical and the
statistics of the observations are assumed to be constant over time,
$\hat{\epsilon}$ given by (\ref{epsilonorig}) is a constant
independent of $k$. From the structure of the expected reward in
(\ref{exprewardnew}) it follows that we can redefine our problem
such that the reward in slot $k$ is now given by,
\begin{eqnarray}
r_k&=& B(1-\hat{\epsilon})\mathcal{I}_{\{S_{u_k}(k) = 0 \}}
\label{rewardnew}
\end{eqnarray}
and the optimization problem is equivalent to maximizing
$\displaystyle \sum_{k=0}^{\infty} \alpha^k \mathsf{E} [r_k]$.
Since we know the structure of the optimal access decisions
from (\ref{access}), the problem of spectrum sensing and access
boils down to choosing the optimal channel to sense in each
slot. Whenever the secondary users sense some channel and make
observations with LLR lower than the threshold, they are free
to access that channel. Thus we have converted the constrained
POMDP problem into an unconstrained POMDP problem as was done
in \cite{chen_zhao_swami}.

\section{Dynamic programming} \label{dp}



The state of the system in slot $k$ denoted by
\[\underline{S}(k) = (S_1(k), S_2(k), \ldots, S_L(k))^\top\] is the
vector of states of the channels in $\mathcal{C}$ that have
independent and identical Markovian evolutions. The channel to
be sensed in slot $k$ is decided in slot $k-1$ and is given by
\[u_k = \mu_k(I_{k-1})\] where $\mu_k$ is a deterministic
function and $I_{k} \triangleq (\underline{X}^{k}, u^{k})$
represents the net information about
past observations and decisions up to slot $k$. 
The reward obtained in slot $k$ is a function of the state in
slot $k$ and $u_k$ as given by (\ref{rewardnew}). We seek the
sequence of channels $\{u_0, u_1, \ldots \}$, such that
$\displaystyle \sum_{k=0}^{\infty} \alpha^k \mathsf{E}[r_k]$ is
maximized. It is easily verified that this problem is a
standard dynamic programming problem with imperfect
observations. It is known \cite{bertsekas} that for such a
POMDP problem, a sufficient statistic at the end of any time
slot $k$, is the probability distribution of the system state
$\underline{S}(k)$, conditioned on all the past observations
and decisions, given by $\mathsf{P}(\{\underline{S}(k) =
\underline{s}\}|I_k)$. Furthermore, since the Markovian
evolution of the different channels are independent of each
other, this conditional probability distribution is
equivalently represented by the set of \emph{beliefs} about the
occupancy states of each channel, i.e., the probability of
occupancy of each channel in slot $k$, conditioned on all the
past observations on channel $a$ and times when channel $a$ was
sensed. We use $p_a(k)$ to represent the belief about channel
$a$ at the end of slot $k$, i.e., $p_a(k)$ is the probability
that the state $S_a(k)$ of channel $a$ in slot $k$ is $1$,
conditioned on all observations and decisions up to time slot
$k$, which is given by
\[p_a(k) = \mathsf{P}(\{S_a(k) = 1\}| \underline{X}_a^k ,K_a^k
).\] We use $\underline{p}(k)$ to denote the $L \times 1$ vector
representing the beliefs about the channels in $\mathcal{C}$.
The initial values of the belief parameters for all channels
are set using the stationary distribution of the Markov chain.
We use $P$ to represent the transition probability matrix for
the state transitions of each channel, with $P(i,j)$
representing the probability that a channel that is in state
$i$ in slot $k$ switches to state $j$ in slot $k+1$. We define,
\begin{eqnarray}
q_a(k) = P(1,1)p_a(k-1) + P(0,1)(1-p_a(k-1)). \label{q}
\end{eqnarray}
This $q_a(k)$ represents the probability of occupancy of channel $a$
in slot $k$, conditioned on the observations up to slot $k-1$. Using
Bayes' rule, the belief values are updated as follows after the
observation in time slot $k$:
\begin{equation}
p_a(k) = \frac{q_a(k) f_1(\underline{X_a}(k))}{q_a(k)
f_1(\underline{X}_a(k)) +(1-q_a(k))f_0(\underline{X}_a(k))}
\label{update1}
\end{equation}
when channel $a$ was selected in slot $k$ (i.e., $u_k = a$),
and $p_a(k) = q_a(k)$ otherwise. Thus from (\ref{update1}) we
see that updates for the sufficient statistic can be performed
using only the joint LLR of the observations,
$\mathcal{L}(\underline{X}_a(k))$, instead of the entire vector
of observations. Furthermore, from (\ref{access}) we also see
that the access decisions also depend only on the LLRs. Hence
we conclude that this problem with vector observations is
equivalent to one with scalar observations where the scalars
represent the joint LLR of the observations of all the
cooperating secondary users. Therefore, in the rest of this
paper, we use a scalar observation model with the observation
made on channel $a$ in slot $k$ represented by $Y_a(k)$. We use
$Y^k$ to denote the set of all observations up to time slot $k$
and $Y_a^k$ to denote the set of all observations on channel
$a$ up to slot $k$.

Hence the new access decisions are given by
\begin{equation}
\delta_a(k) = \mathcal{I}_{\{\mathcal{L'}(Y_a(k)) <
\Delta'\}}\mathcal{I}_{\{u_k = a \} } 
\end{equation}
where $\mathcal{L'}(Y_a(k))$ represents the LLR of $Y_a(k)$ and
the access threshold $\Delta'$ is chosen to satisfy,
\begin{equation}
\mathsf{P}(\{\mathcal{L'}(Y_a(k)) < \Delta'\} | \{S_a(k) = 1\}) =
\zeta. \label{tau2}
\end{equation}
Similarly the belief updates are performed as in (\ref{update1})
with the evaluations of density functions of $\underline{X}_a(k)$
replaced with the evaluations of the density functions $f_0'$ and
$f_1'$ of $Y_a(k)$:
\begin{equation}
p_a(k) = \frac{q_a(k) f_1'(Y_a(k))}{q_a(k) f_1'(Y_a(k))
+(1-q_a(k))f_0'(Y_a(k))} \label{update2}
\end{equation}
when channel $a$ is accessed in slot $k$ (i.e., $u_k = a$), and
$p_a(k) = q_a(k)$ otherwise. We use $G(\underline{p}(k-1),u_k,
Y_{u_k}(k))$ to denote the function that returns the value of
$\underline{p}(k)$ given that channel $u_k$ was sensed in slot $k$.
This function can be calculated using the relations (\ref{q}) and
(\ref{update2}).
The reward obtained in slot $k$ can now be expressed as,
\begin{eqnarray}
r_k&=& B(1-\epsilon)\mathcal{I}_{\{S_{u_k}(k) = 0 \}} \label{reward}
\end{eqnarray}
where $\epsilon$ is given by
\begin{equation}
\epsilon = \mathsf{P}(\{\mathcal{L'}(Y_a(k)) > \Delta'\}| \{S_a(k) =
0\}). \label{epsilon}
\end{equation}

From the structure of the dynamic program, it can be shown 
that the optimal solution to this dynamic program can be obtained by
solving the following Bellman equation \cite{bertsekas} for the
optimal reward-to-go function:
\begin{eqnarray}
J(\underline{p}) = \max_{u \in \mathcal{C}}[B(1-\epsilon)(1-q_{u}) + \alpha
\mathsf{E}(J(G(\underline{p},u,Y_{u})))] \label{bellman}
\end{eqnarray}
where $\underline{p}$ represents the initial value of the belief
vector, i.e., the prior probability of channel occupancies in slot
$-1$, and $\underline{q}$ is calculated from $\underline{p}$ as in
(\ref{q}) by,
\begin{eqnarray}
q_a = P(1,1)p_a + P(0,1)(1-p_a) , \hspace{0.5 in}a \in \mathcal{C}.\label{q2}
\end{eqnarray}
The expectation in (\ref{bellman}) is performed over the random
observation $Y_{u}$. Since it is not easy to find the optimal
solution to this Bellman equation, we adopt a suboptimal strategy to
obtain a channel selection policy that performs well.


In the rest of the paper we assume that the transition
probability matrix $P$ satisfies the following regularity
conditions:
\begin{eqnarray}
\mbox{Assumption 1}&:& 0 < P(j,j) < 1, \quad j \in \{0,1 \} \label{assnew}\\
\mbox{Assumption 2}&:& P(0,0) > P(1,0) \label{ass1}
\end{eqnarray}
The first assumption ensures that the resultant Markov
chain is irreducible and positive recurrent, while the second assumption ensures that it is
more likely for a channel that is free in the current slot to remain free in the next slot than
for a channel that is occupied in the current slot to switch states and become free in the next
slot. While the first assumption is important the second one is used only in the derivation of
the upper bound on the optimal performance and can easily be relaxed by separately considering
the case where the inequality (\ref{ass1}) does not hold.
\subsection{Greedy policy} \label{greedypolicy}
A straightforward suboptimal solution to the channel selection
problem is the greedy policy, i.e., the policy of maximizing
the expected instantaneous reward in the current time slot. The
expected instantaneous reward obtained by accessing some
channel $a$ in a given slot $k$ is given by $B(1-\epsilon)(1-
q_a(k))$ where $\epsilon$ is given by (\ref{epsilon}). Hence
the greedy policy is to choose the channel $a$ such that $1 -
q_a(k)$ is the maximum.
\begin{equation}
u_k^{\sf{gr}} = \underset{u \in \mathcal{C}}{\operatorname{argmax}}\,\{1 - q_u(k) \}.
\end{equation}
In other words, in every slot the greedy policy chooses the
channel that is most likely to be free, conditioned on the past
observations. The greedy policy for this problem is in fact
equivalent to the $Q_{\text{MDP}}$ policy, which is a standard
suboptimal solution to the POMDP problem (see, e.g.,
\cite{aberdeen}). It is shown in \cite{zhao_krish_liu} and
\cite{javidi_krish_zhao_liu} that under some conditions on $P$
and $L$, the greedy policy is optimal if the observation in
each slot reveals the underlying state of the channel. Hence it
can be argued that under the same conditions, the greedy policy
would also be optimal for our problem at high SNR.

\subsection{An upper bound} \label{upperbound}
An upper bound on the optimal reward for the POMDP of
(\ref{bellman}) can be obtained by assuming more information
than the maximum that can be obtained in reality. One such
assumption that can give us a simple upper bound is the
$Q_{\text{MDP}}$ assumption \cite{aberdeen},
which is to assume that in all future slots, the state of all
channels become known exactly after making the observation in
that slot. The optimal reward under the $Q_{\text{MDP}}$
assumption is a function of the initial belief vector, i.e.,
the prior probabilities of occupancy of the channels in slot
$-1$. We represent this function by $J^Q$. In practice, a
reasonable choice of initial value of the belief vector is
given by the stationary distribution of the Markov chains.
Hence for any solution to the POMDP that uses this
initialization, an upper bound for the optimal reward under the
$Q_{\text{MDP}}$ assumption is given by $J^U =
J^Q(p^*\underline{1})$ where $p^*$ represents the probability
that a channel is occupied under the stationary distribution of
the transition probability matrix $P$, and $\underline{1}$
represents an $L \times 1$ vector of all $1$'s.

The first step involved in evaluating this upper bound is to
determine the optimal reward function under the assumption that all
the channel states become known exactly after making the observation
in each slot including the current slot. We call this function
$\tilde{J}$. That is, we want to evaluate $\tilde{J}(\underline{x})$
for all binary strings $\underline{x}$ of length $L$ that represent
the $2^L$ possible values of the vector representing the states of
all channels in slot $-1$. The $Q_{\text{MDP}}$ assumption implies
that the functions $J^Q$ and $\tilde{J}$ satisfy the following
equation:
\begin{eqnarray}
J^Q(\underline{z}) &=& \max_{u \in \mathcal{C}}\{[B(1-\epsilon)(1-q_{u}) + \qquad \qquad \qquad \qquad \nonumber\\
&& \qquad \qquad  \sum_{\underline{x}
\in \{0,1 \}^L} \alpha \mathsf{P}(\{\underline{S}(0) = \underline{x}\})
\tilde{J}(\underline{x})] \} \mbox{ s.t. } \underline{p}(-1) = \underline{z}
\label{qmdpcostfn}
\end{eqnarray}
where $\underline{p}(-1)$ denotes the a priori belief vector
about the channel states in slot $-1$ and $q_u$ is obtained
from $p_u(-1)$ just as in (\ref{q2}). Hence the upper bound
$J^U = J^Q(p^*\underline{1})$ can be easily evaluated using $P$
once the function $\tilde{J}$ is determined.


Now we describe how one can solve for the function $\tilde{J}$. Under the assumption that the
states of all the channels become known exactly at the time of observation, the optimal channel
selected in any slot $k$ would be a function of the states of the channels in slot $k-1$.
Moreover, the sensing action in the current slot would not
affect the rewards in the future slots. Hence the optimal
policy would be to maximize the expected instantaneous reward,
which is achieved by accessing the channel that is most likely
to be free in the current slot. Now under the added assumption
stated in (\ref{ass1}) earlier\footnote{It is easy to see that
a minor modification of the derivation of the upper bound works
when assumption (\ref{ass1}) does not hold.},
the optimal policy would always select some channel that was
free in the previous time slot, if there is any. If no channel
is free in the previous time slot, then the optimal policy
would be to select any one of the channels in $\mathcal{C}$,
since all of them are equally likely to be free in the current
slot. Hence the derivation of the optimal total reward for this
problem is straightforward as illustrated below. The total
reward for this policy is a function of the state of the system
in the slot preceding the initial slot, i.e.,
$\underline{S}(-1)$.
\begin{eqnarray}
\tilde{J}(\underline{x}) &=& \max_{u \in \mathcal{C}}\mathsf{E} \left[\kappa[\mathcal{I}_{\{S_{u}(0) =
0\}} + \alpha \tilde{J}(\underline{S}(0))]\bigg{|}\{\underline{S}(-1)= \underline{x}\} \right]\nonumber \\
&=& \left\{ \begin{tabular}{c} $\kappa P(0,0) + \alpha
V(\underline{x})$ if $\underline{x} \neq \underline{1}$\\
$\kappa P(1,0) + \alpha V(\underline{x})$ if $\underline{x} =
\underline{1}$
\end{tabular} \right. \nonumber
\end{eqnarray}
where $V(\underline{x}) =
\mathsf{E}[\tilde{J}(\underline{S}(0))|\{\underline{S}(-1)=
\underline{x}\}]$, $\kappa = B (1-\epsilon)$, and $
\underline{1}$ is an $L \times 1$ string of all $1$'s. This
means that we can write
\begin{eqnarray}
\tilde{J}(\underline{x})&=& \kappa \left\{ P(0,0) \sum_{k =
0}^\infty \alpha^k - (P(0,0) - P(1,0) )w(\underline{x})\right\}
\nonumber\\
&=& \kappa \left\{ \frac{P(0,0)}{1-\alpha} - (P(0,0) - P(1,0)
)w(\underline{x}) \right\} \label{jbound}
\end{eqnarray}
where \[ w(\underline{x}) \triangleq \mathsf{E}\left[\displaystyle\sum_{M
\geq -1: \underline{S}(M) = \underline{1}} \alpha^{M+1} \Bigg{|}
\{\underline{S}(-1)=\underline{x}\} \right]\] is a scalar function
of the vector state $\underline{x}$. Here the expectation is over
the random slots when the system reaches state $\underline{1}$. Now
by stationarity we have,
\begin{equation}
w(\underline{x}) = \mathsf{E}\left[\sum_{M \geq 0: \underline{S}(M)
= \underline{1}} \alpha^{M} \Bigg{|} \{
\underline{S}(0)=\underline{x}\} \right]. \label{weqn}
\end{equation}

We use $\mathbb{P}$ to denote the matrix of size $2^L \times 2^L$
representing the transition probability matrix of the joint Markov
process that describes the transitions of the vector of channel
states $\underline{S}(k)$. The $(i,j)^\text{th}$ element of
$\mathbb{P}$ represents the probability that the state of the system
switches to $\underline{y}$ in slot $k+1$ given that the state of
the system is $\underline{x}$ in slot $k$, where $\underline{x}$ is
the $L$-bit binary representation of $i-1$ and $\underline{y}$ is
the $L$-bit binary representation of $j-1$. Using a slight abuse of
notation we represent the $(i,j)^\text{th}$ element of $\mathbb{P}$
as $\mathbb{P}(\underline{x},\underline{y})$ itself. Now equation
(\ref{weqn}) can be solved to obtain,
\begin{equation}
w(\underline{x}) = \sum_{\underline{y}} \alpha
\mathbb{P}(\underline{x},\underline{y})w(\underline{y}) +
\mathcal{I}_{\{\underline{x} = \underline{1} \}}.
\end{equation}
This fixed point equation which can be solved to obtain,
\begin{equation}
\underline{w} = (\mathbb{I} - \alpha \mathbb{P})^{-1}\left(
\begin{tabular}{c}$0$\\$\vdots$\\$0$\\$1$\end{tabular} \right)_{2^L \times
1} \label{wfixedpoint}
\end{equation}
where $\underline{w}$ is a $2^L \times 1$ vector whose elements are
the values of the function $w(\underline{x})$ evaluated at the $2^L$
different possible values of the vector state $\underline{x}$ of the
system in time slot $-1$. Again, the $i^\text{th}$ element of vector
$\underline{w}$ is $w(\underline{x})$ where $\underline{x}$ is the
$L$-bit binary representation of $i-1$. Thus $\tilde{J}$ can now be
evaluated by using relation (\ref{jbound}) and the expected reward
for this problem under the $Q_{\text{MDP}}$ assumption can be
calculated by evaluating $J^U = J^Q(p^*\underline{1})$ via equation
(\ref{qmdpcostfn}). This optimal value yields an analytical upper
bound on the optimal reward of the original problem (\ref{bellman}).


\subsection{Comparison with \cite{chen_zhao_swami} for single user problem} \label{comparison}
Although we have studied a spectrum access scheme for a
cooperative cognitive radio network, it can also be employed by
a single cognitive user. Under this setting, our approach to
the spectrum access problem described earlier in this section
is similar to that considered in \cite{chen_zhao_swami} and
\cite{zhao_tong_swami} in that sensing does not reveal the true
channel states but only a random variable whose distribution
depends on the current state of the sensed channel. As a
result, the structure of our optimal access policy and the
sufficient statistic are similar to those in
\cite{chen_zhao_swami}. In this section we compare the two
schemes.

The main difference between our formulation and that in
\cite{chen_zhao_swami} is that in our formulation the secondary
users use the primary signal received on the channel to track
the channel occupancies, while in \cite{chen_zhao_swami} they
use the ACK signals exchanged between the secondary transmitter
and receiver. Under the scheme of \cite{chen_zhao_swami}, in
each slot, the secondary receiver transmits an ACK signal upon
successful reception of a transmission from the secondary
receiver. The belief updates are then performed using the
single bit of information provided by the presence or absence
of the ACK signal. The approach of \cite{chen_zhao_swami} was
motivated by the fact that, under that scheme, the secondary
receiver knows in advance the channel on which to expect
potential transmissions from the secondary transmitter in each
slot, thus obviating the need for control channels for
synchronization. However, such synchronization between the
transmitter and receiver is not reliable in the presence of
interfering terminals that are hidden \cite{jstsp} from either
the receiver or transmitter, because the ACK signals will no
longer be error-free. In this regard, we believe that a more
practical solution to this problem would be to set aside a
dedicated control channel of low capacity for the purpose of
reliably maintaining synchronization, and use the observations
on the primary channel for tracking the channel occupancies. In
addition to guaranteeing synchronization, our scheme provides
some improvement in utilizing transmission opportunities over
the ACK-based scheme, as we show in section \ref{sim1}.

Another difference between our formulation and that in
\cite{chen_zhao_swami} is that we assume that the statistics of
channel occupancies are independent and identical while
\cite{chen_zhao_swami} considers the more general case of
correlated and non-identical channels. However, the scheme we
proposed in section \ref{dp} can be easily modified to handle
this case, with added complexity. The sufficient statistic
would now be the posteriori distribution of $\underline{S}(k)$,
the vector of states of all channels, and the access thresholds
on different channels would be non-identical and depend on the
statistics of the observations the respective channels. We
avoid elaborating on this more general setting to keep the
presentation simple.


\section{The case of unknown distributions} \label{unknown}
In practice, the secondary users are typically unaware of the
primary's signal characteristics and the channel realization
from the primary \cite{somefund}. Hence cognitive radio systems
have to rely on some form of non-coherent detection such as
energy detection while sensing the primary signals.
Furthermore, even while employing non-coherent detectors, the
secondary users are also unaware of their locations relative to
the primary and hence are not aware of the shadowing and path
loss from the primary to the secondary. Hence it is not
reasonable to assume that the secondary users know the exact
distributions of the observations under the primary-present
hypothesis, although it can be assumed that the distribution of
the observations under the primary-absent hypothesis is known
exactly. This scenario can be modeled by using a parametric
description for the distributions of the received signal under
the primary-present hypothesis. We denote the density functions
of the observations under the two possible hypotheses as,
\begin{eqnarray}
S_a(k) = 0&:& {{Y_a(k)}} \sim f_{\theta_0} \nonumber \\
S_a(k) = 1&:& {{Y_a(k)}} \sim f_{\theta_a} \nonumber \\
\mbox{ where } \theta_a \in \Theta, \forall a
&\in& \{1,
2, \ldots, L \} \label{parameter}
\end{eqnarray}
where the parameters $\{\theta_a\}$ are unknown for all channels
$a$, and $\theta_0$ is known. We use
$\mathcal{L}_{\theta}(.)$ to denote the log-likelihood function
under $f_{\theta}$ defined by,
\begin{equation}
\mathcal{L}_{\theta}(x) \triangleq \log \left(
\frac{f_{\theta}(x)}{f_{\theta_0}(x)}\right), \hspace{0.5 in} x \in
\mathbb{R}, \theta \in \Theta. \label{llr}
\end{equation}


In this section, we study two possible approaches for dealing
with such a scenario, while restricting to greedy policies for
channel selection. For ease of illustration, in this section we
consider a secondary system comprised of a single user,
although the same ideas can also be applied for a system with
multiple cooperating users.

\subsection{Worst-case design for non-random $\theta_a$} \label{worstcase}
A close examination of Section \ref{dp} reveals two specific
uses for the density function of the observations under the
$S_a(k) = 1$ hypothesis. The knowledge of this density was of
crucial importance in setting the access threshold in
(\ref{tau2}) to meet the constraint on the probability of
interference. The other place where this density was used was
in updating the belief probabilities in (\ref{update2}). When
the parameters $\{\theta_a \}$ are non-random and unknown, we
have to guarantee the constraint on the interference
probability for all possible realizations of $\theta_a$. The
optimal access decision would thus be given by,
\begin{equation}
\hat \delta_a(k) = \mathcal{I}_{\{u_k = a \}} \prod_{\theta \in
\Theta} \mathcal{I}_{\{\mathcal{L}_{\theta}(Y_a(k)) <
\tau_{\theta}\}} \label{accessworst}
\end{equation}
where $\tau_{\theta}$ satisfies,
\begin{equation}
\mathsf{P}(\{\mathcal{L}_{\theta}(Y_a(k)) < \tau_{\theta}\} |
\{S_a(k) = 1, \theta_a = \theta \}) = \zeta. \label{tau3}
\end{equation}
The other concern that we need to address in this approach is: what
distribution do we use for the observations under $S_a(k) = 1$ in
order to perform the updates in (\ref{update2}). An intelligent solution
is possible provided
the densities described in (\ref{parameter}) satisfy the condition
that there is a $\theta^* \in \Theta$ such that the for all $\theta
\in \Theta$ and for all $\tau \in \mathbb{R}$ the following
inequality holds:
\begin{eqnarray}
\mathsf{P}(\{\mathcal{L}_{\theta^*}(Y_a(k)) > \tau\}|\{S_a(k) = 1,
\theta_a = \theta \}) \geq \nonumber \\
\mathsf{P}(\{\mathcal{L}_{\theta^*}(Y_a(k)) > \tau\}|\{S_a(k) = 1,
\theta_a = \theta^* \}). \label{condition}
\end{eqnarray}
The condition (\ref{condition}) is satisfied by several
parameterized densities including an important practical
example discussed later. Under condition (\ref{condition}), a
good suboptimal solution to the channel selection problem would
be to run the greedy policy for channel selection using
$f_{\theta^*}$ for the density under $S_a(k) = 1$ while
performing the updates of the channel beliefs in
(\ref{update2}). This is a consequence of the following lemma.


\begin{lem}\label{lemma}
Assume condition (\ref{condition}) holds. Suppose $f_\theta^*$ is used in place of $f_1'$ for the distribution of the observations
under $S_a(k) = 1$ while performing belief updates in (\ref{update2}). Then,
\begin{romannum}
\item For all $\gamma \in \Theta$ and for all $\beta, p \in [0,1]$,
\begin{eqnarray}
\mathsf{P}_{\gamma} (\{p_a(k) > \beta\}| \{S_a(k) =1,    p_a(k-1) = p\} )  \geq \nonumber \\
\mathsf{P}_{\theta^*} (\{p_a(k) > \beta\}| \{S_a(k) =1,    p_a(k-1) = p\} ) \label{belf_stoch_order}
\end{eqnarray}
where $\mathsf{P}_\theta$ represents the probability measure when $\theta_a = \theta$.
\item Conditioned on $\{S_a(k) =0\}$, the distribution of $p_a(k)$ given any value for $p_a(k-1)$ is identical for all possible values of $\theta_a$.
\end{romannum}
\end{lem}
\begin{proof}
(i) Clearly (\ref{belf_stoch_order}) holds with equality when channel $a$ is not sensed in slot $k$ (i.e. $u_k \neq a$). When $u_k = a$, it is easy to
see that the new belief given by (\ref{update2}) is a monotonically increasing function of the log-likelihood function, $\mathcal{L}_{\theta^*}(Y_a(k))$. Hence
(\ref{belf_stoch_order}) follows from condition (\ref{condition}).

(ii) This is obvious since the randomness in $p_a(k)$ under $\{S_a(k) =0\}$ is solely due to the observation $Y_a(k)$ whose distribution $f_{\theta_0}$ does not depend on $\theta_a$.
\end{proof}

Clearly, updating using $f_{\theta^*}$ in (\ref{update2}) is optimal if $\theta_a = \theta^*$. When $\theta_a \neq \theta^*$,
the tracking of beliefs are guaranteed to be at least as accurate, in the sense described in Lemma \ref{lemma}.
Hence, under condition (\ref{condition}), a good suboptimal
solution to the channel selection problem would be to run the
greedy policy for channel selection using $f_{\theta^*}$ for
the density under $S_a(k) = 1$ while performing the updates of
the channel beliefs in (\ref{update2}). Furthermore, it is
known that \cite{venu_basar_poor} under condition
(\ref{condition}), the set of likelihood ratio tests in the
access decision of (\ref{accessworst}) can be replaced with a
single likelihood ratio test under the worst case parameter
$\theta^*$ given by,
\begin{equation}
\hat \delta_a(k) = \mathcal{I}_{\{u_k = a \}}
\mathcal{I}_{\{\mathcal{L}_{\theta^*}(Y_a(k)) < \tau_{\theta^*}\}  }.
\label{accessworstnew}
\end{equation}
The structure of the access decision given in (\ref{accessworstnew}), and the conclusion from Lemma \ref{lemma}
suggests that $\theta^*$ is a worst-case value of the parameter $\theta_a$. Hence the strategy of designing
the sensing and access policies assuming this worst possible value of the parameter is optimal in the following min-max sense:
The average reward when the true value of $\theta_a \neq \theta^*$ is expected to be no smaller than that obtained when
$\theta_a = \theta^*$ since the tracking of beliefs is worst when $\theta_a = \theta^*$ as shown in Lemma \ref{lemma}. This
intuitive reasoning is seen to hold in the simulation results in Section \ref{sim2}.

\subsection{Modeling $\theta_a$ as random} \label{modeltheta}
%
%

In Section \ref{sim2}, we show through simulations that the
worst-case approach of the previous section leads to a severe
decline in performance relative to the scenario where the
distribution parameters in (\ref{parameter}) are known
accurately. In practice it may be possible to learn the value
of these parameters online. In order to learn the parameters
$\{\theta_a\}$ we need to have a statistical model for these
parameters and a reliable statistical model for the channel
state process. In this section we model the parameters
$\{\theta_a\}$ as random variables, which are i.i.d. across the
channels and independent of the Markov process as well as the
noise process. In order to assure the convergence of our
learning algorithm, we also assume that the cardinality of set
$\Theta$ is finite\footnote{We do discuss the scenario when
$\Theta$ is a compact set in the example considered in Section
\ref{sim2}.} and let $|\Theta| = N$. Let $\{\mu_i\}_1^N$ denote
the elements of set $\Theta$. The prior distribution of the
parameters $\{\theta_a\}$ is known to the secondary users. The
beliefs of the different channels no longer form a sufficient
statistic for this problem. Instead, we keep track of the
following set of a posteriori probabilities which we refer to
as \emph{joint beliefs}:
\begin{eqnarray}
\left\{\mathsf{P}(\{(\theta_a, S_a(k)) = (\mu_i,j)\}|I_k) :
\forall i, j, a\right\}.
\end{eqnarray}
Since we assume that the parameters $\{\theta_a\}$ take values
in a finite set, we can keep track of these joint beliefs just
as we kept track of the beliefs of the states of different
channels in Section \ref{dp}. For the initial values of these
joint beliefs we use the product distribution of the stationary
distribution of the Markov chain and the prior distribution on
the parameters $\{\theta_a\}$. We store these joint beliefs at
the end of slot $k$ in an $L \times N \times 2$ array $Q(k)$
with elements given by,
\begin{equation}
Q_{a,i, j}(k) = \mathsf{P}(\{(\theta_a, S_a(k)) = (\mu_i,j)
\}|Y_a^k, K_a^k). \label{Qmatrix}
\end{equation}
The entries of the array $Q(k)$ corresponding to channel $a$
represent the joint a posteriori probability distribution of the
parameter $\theta_a$ and the state of channel $a$ in slot $k$
conditioned on the information available up to slot $k$ which we
called $I_k$. Now define,
\[H_{a,i,j}(k) = \displaystyle \sum_{\ell \in \{0, 1\}} P(\ell,j)Q_{a,i,\ell}(k-1).\]
Again, the values of the array $H(k)$ represent the a posteriori
probability distributions about the parameters $\{\theta_a\}$ and
the channel states in slot $k$ conditioned on $I_{k-1}$, the
information up to slot $k-1$. The update equations for the joint
beliefs can now be written as follows:
\[
Q_{a,i,j}(k) =\left\{ \begin{tabular}{c} $\lambda{
H_{a,i,0}(k)f_{\theta_0}(Y_a(k))} $ if $j = 0$\\
$\lambda {H_{a,i,1}(k)f_{\mu_{i}}(Y_a(k))} $ if $j = 1$
\end{tabular} \right. \nonumber
\]
when channel $a$ was accessed in slot $k$, and $Q_{a,i,j}(k) =
H_{a,i,j}(k)$ otherwise. Here $\lambda$ is just a normalizing
factor.

It is shown in Appendix that, for each channel $a$, the a
posteriori probability mass function of parameter $\theta_a$
conditioned on the information up to slot $k$, converges to a
delta-function at the true value of parameter $\theta_a$ as $k
\to \infty$, provided we sense channel $a$ frequently enough.
This essentially means that we can learn the value of the
actual realization of $\theta_a$ by just updating the joint
beliefs. This observation suggests that we could use this
knowledge learned about the parameters in order to obtain
better performance than that obtained under the policy of
Section \ref{worstcase}. We could, for instance, use the
knowledge of the true value of $\theta_a$ to be more liberal in
our access policy than the satisfy-all-constraints approach
that we used in Section \ref{worstcase} when we did a
worst-case design. With this in mind, we propose the following
algorithm for choosing the threshold to be used in each slot
for determining whether or not to access the spectrum.

Assume channel $a$ was sensed in slot $k$. We first arrange the
elements of set $\Theta$ in increasing order of the a posteriori
probabilities of parameter $\theta_a$. We partition $\Theta$ into
two groups, a `lower' partition and an `upper' partition, such that
all elements in the lower partition have lower a posteriori
probability values than all elements in the upper partition. The
partitioning is done such that the number of elements in the lower
partition is maximized subject to the constraint that the a
posteriori probabilities of the elements in the lower partition add
up to a value lower than $\zeta$. These elements of $\Theta$ can be
ignored while designing the access policy since the sum of their a
posteriori probabilities is below the interference constraint. We
then design the access policy such that we meet the interference
constraint conditioned on parameter $\theta_a$ taking any value in
the upper partition. The mathematical description of the algorithm
is as follows. Define
\[b_a^i(k) \triangleq \displaystyle \sum_{j \in \{0, 1\}} Q_{a,i,j}(k-1).\]
The vector $(b_a^1(k), b_a^2(k), \ldots, b_a^N(k))^\top$ represents the a posteriori
probability mass function of parameter $\theta_a$ conditioned on $I_{k-1}$, the information
available up to slot $k-1$. Now let $\pi_k(i): \{1, 2, \ldots, N\} \mapsto \{1, 2, \ldots, N\}$
be a permutation of $\{1, 2, \ldots, N\}$ such that $\{\mu_{\pi_k(i)}\}_{i = 1}^N$ are arranged
in increasing order of posteriori probabilities, i.e. \[i \geq j \Leftrightarrow
b_a^{\pi_k(i)}(k) \geq b_a^{\pi_k(j)}(k)\] and let $N_a(k) = \displaystyle \max \{c \leq N :
\sum_{i = 1}^c b_a^{\pi_k(i)}(k) < \zeta\}$. Now define set $\Theta_a(k) = \{\mu_{\pi_k(i)}: i
\geq N_a(k)\}$. This set is the upper partition mentioned earlier. The access decision on
channel $a$ in slot $k$ is given by,\footnote{The access policy obtained via the partitioning
scheme is simple to implement but is not the optimal policy in general. The optimal access
decision on channel $a$ in slot $k$ would be given by a likelihood-ratio test between
$f_{\theta_0}$ and the mixture density $\sum_{\theta \in \Theta} r_\theta(k-1) f_\theta$ where
$r_\theta(k-1)$ represents the value of the posterior distribution of $\theta_a$ after slot
$k-1$, evaluated at $\theta$. However setting thresholds for such a test is prohibitively
complex.}
\begin{equation}
\tilde \delta_a(k) = \mathcal{I}_{\{u_k = a \}} \prod_{\theta \in
\Theta_a(k)} \mathcal{I}_{\{\mathcal{L}_{\theta}(Y_a(k)) <
\tau_{\theta}\}} \label{accessrandom}
\end{equation}
where $\tau_{\theta}$ satisfy (\ref{tau3}).
The access policy given above guarantees that
\begin{equation}
\mathsf{P}(\{\tilde \delta_a(k) = 1\}| \{S_a(k) = 1\}, Y^{k-1},
K^{k-1}) < \zeta \label{thetaguarantee}
\end{equation} whence the same holds without conditioning on
$Y^{k-1}$ and $K^{k-1}$. Hence, the interference constraint is met
on an average, averaged over the posteriori distributions of
$\theta_a$. Now it is shown in Appendix that the a posteriori
probability mass function of parameter $\theta_a$ converges to a
delta function at the true value of parameter $\theta_a$ almost
surely. Hence the constraint is asymptotically met even conditioned
on $\theta_a$ taking the correct value. This follows from the fact
that, if $\mu_{i*}$ is the actual realization of the random variable
$\theta_a$, and $b_a^{i*}(k)$ converges to $1$ almost surely, then,
for sufficiently large $k$, (\ref{accessrandom}) becomes: $\tilde
\delta_a(k) = \mathcal{I}_{\{u_k = a \}}
\mathcal{I}_{\{\mathcal{L}_{\mu_{i*}}(Y_a(k)) < \tau_{\mu_{i*}}\} }$
with probability one and hence the claim is satisfied.

It is important to note that the access policy given in
(\ref{accessrandom}) need not be the optimal access policy for this
problem. Unlike in Section \ref{probform}, here we are allowing the
access decision in slot $k$ to depend on the observations in all
slots up to $k$ via the joint beliefs. Hence, it is no longer
obvious that the optimal test should be a threshold test on the
LLR of the observations in the current slot even if
parameter $\theta_a$ is known. However, this structure for the
access policy can be justified from the observation that it is
simpler to implement in practice than some other policy that
requires us to keep track of all the past observations. The
simulation results that we present in Section \ref{sim2} also
suggest that this scheme achieves substantial improvement in
performance over the worst-case approach, thus further justifying
this structure for the access policy.



%

Under this scheme the new greedy policy for channel selection is to
sense the channel which promises the highest expected instantaneous
reward which is now given by,

\begin{equation}
\widetilde{u_k^{\sf{gr}}} = \underset{a \in \mathcal{C}}{\operatorname{argmax}}\, \left\{\displaystyle \sum_{i =1}^N
h_{a,i,0}(k)(1-\epsilon_a(k)) \right\} \label{selectionrandom}
\end{equation}
where
\[\epsilon_a(k) = \mathsf{P}\left(\bigcup_{\theta \in \Theta_a(k)}\{\mathcal{L}_{\theta}(Y_a(k))
> \tau_{\theta}\}\bigg{|} \{S_a(k) = 0\}\right).\]
However, in order to prove the convergence of the a posteriori
probabilities of the parameters $\{\theta_a\}$, we need to make a
slight modification to this channel selection policy. In our proof,
we require that each channel is accessed frequently. To enforce that
this condition is satisfied, we modify the channel selection policy
so that the new channel selection scheme is as follows:
\begin{eqnarray}
\widetilde{u_k^{\sf{mod}}} = \left\{ \begin{tabular}{cc}
$\mathcal{C}_j$ & if $k \equiv j\mod CL, \hspace{0.25 in} j \in \mathcal{C}$\\
$\widetilde{u_k^{\sf{gr}}}$ &else
\end{tabular} \right.  \label{selectionrandommod}
\end{eqnarray}
where $C > 1$ is some constant and $\{\mathcal{C}_j: 1\leq j
\leq L\}$ is some ordering of the channels in $\mathcal{C}$.

%


\section{Simulation results and comparisons} \label{sims}

\subsection{Known distributions} \label{sim1}

We consider a simple model for the distributions of the
observations and illustrate the advantage of our proposed
scheme over that in \cite{chen_zhao_swami} by simulating the
performances obtained by employing the greedy algorithm on both
these schemes. We also consider a combined scheme that uses
both the channel observations and the ACK signals for updating
beliefs.


We simulated the greedy policy under three different schemes.
Our scheme, which we call $G_1$, uses only the observations
made on the channels to update the belief vectors. The second
one, $G_2$, uses only the ACK signals transmitted by the
secondary receiver, while the third one, $G_3$, uses both
observations as well as the error-free ACK signals. We have
performed the simulations for two different values of the
interference constraint $\zeta$. The number of channels was
kept at $L=2$ in both cases and the transition probability
matrix used was,
\[P = \left[
\begin{array}{cc} 0.9 & 0.1\\0.2 & 0.8
\end{array} \right]\] where the first index represents state 0 and
the second represents state 1. Both channels were assumed to
have unit bandwidth, $B=1$ and the discount factor was set to
$\alpha = 0.999$. Such a high value of $\alpha$ was chosen to
approximate the problem with no discounts which would be the
problem of practical interest. As we saw in Section \ref{dp},
the spectrum access problem with a group of cooperating
secondary users is equivalent to that with a single user.
Hence, in our simulations we use a scalar observation model
with the following simple distributions for $Y_a(k)$ under the
two hypotheses:
\begin{eqnarray}
S_a(k) = 0 \mbox{ (primary OFF)}&:& {Y_a(k)} \sim \mathcal{N}(0,
\sigma^2) \nonumber \\
S_a(k) = 1 \mbox{ (primary ON)} &:& {Y_a(k)} \sim \mathcal{N}(\mu,
\sigma^2) \label{hypotheses}
\end{eqnarray}
It is easy to verify that the LLR for these observations is an
increasing linear function of $Y_a(k)$. Hence the new access
decisions are made by comparing $Y_a(k)$ to a threshold $\tau$
chosen such that,
\begin{equation}
\mathsf{P}(\{Y_a(k) < \tau\} | \{S_a(k) = 1\}) = \zeta \label{tau}
\end{equation}
and access decisions are given by,
\begin{equation}
\delta_a(k) = \mathcal{I}_{\{Y_a(k) < \tau\}} \mathcal{I}_{\{u_k = a
\} }.
\end{equation}
The belief updates in (\ref{update1}) are now given by,
\begin{eqnarray*}
p_a(k) = \frac{q_a(k) f(\mu,\sigma^2, Y_a(k))}{q_a(k)
f(\mu,\sigma^2, Y_a(k)) +(1-q_a(k))f(0,\sigma^2, Y_a(k))}
\end{eqnarray*}
when channel $a$ was selected in slot $k$ (i.e. $u_k = a$), and
$p_a(k) = q_a(k)$ otherwise. Here $q_a(k)$ is given by
(\ref{q}) and $f(x,y,z)$ represents the value of the Gaussian
density function with mean $x$ and variance $y$ evaluated at
$z$. For the mean and variance parameters in (\ref{hypotheses})
we use $\sigma = 1$ and choose $\mu$ so that $\sf{SNR} = 20
\log_{10} (\mu/\sigma)$ takes values from $-5$ dB to $5$ dB. In
the case of cooperative sensing, this $\sf{SNR}$ represents the
effective signal-to-noise ratio in the joint LLR statistic at
the decision center, $\mathcal{L}(\underline{X}_a(k))$. We
perform simulations for two values of the interference
constraint, $\zeta = 0.1$ and $\zeta = 0.01$.



\begin{figure}
\centering
\includegraphics[width=7in]{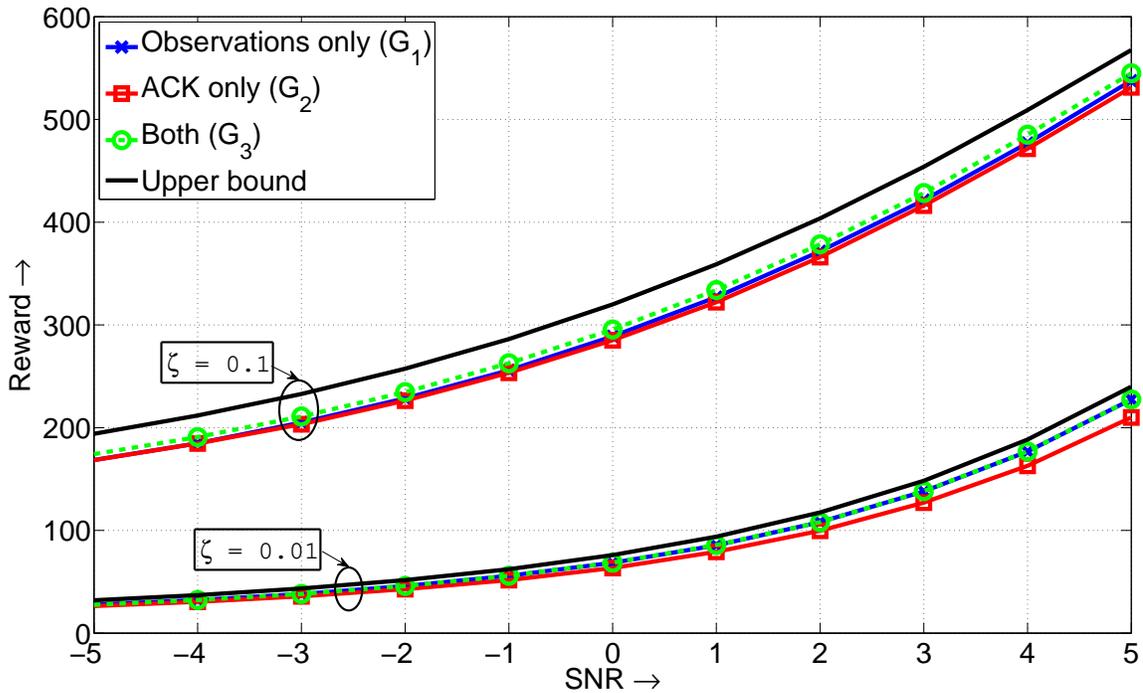}
\caption{Comparison of performances obtained with greedy policy that
uses observations and greedy policy that uses ACKs.
Performance obtained with greedy policy that uses both ACKs as well
as observations and the upper bound are also shown.}
\figlbl{figsim1}
\end{figure}


As seen in \figref{figsim1}, the strategy of using only ACK
signals ($G_2$) performs worse than the one that uses all the
observations ($G_1$), especially for $\zeta = 0.01$, thus
demonstrating that relying only on ACK signals compromises on
the amount of information that can be learned. We also observe
that the greedy policy attains a performance that is within $10
\%$ of the upper bound.
It is also seen in the figure that the reward values obtained
under $G_1$ and $G_3$ are almost equal. For $\zeta =0.01$, it
is seen that the two curves are overlapping. This observation
suggests that the extra advantage obtained by incorporating the
ACK signals is insignificant especially when the interference
constraint is low. The explanation for this observation is that
the ACK signals are received only when the signal transmitted
by the secondary transmitter successfully gets across to its
receiver. For this to happen the state of the primary channel
should be `0' and the secondary must decide to access the
channel. When the value of the interference constraint $\zeta$
is low, the secondary accesses the channel only when the value
of the $Y_a(k)$ is low. Hence the observations in this case
carry a significant amount of information about the states
themselves and the additional information that can be obtained
from the ACK signals is not significant. Thus learning using
only observations is almost as good as learning using both
observations as well as ACK signals in this case.

\subsection{Unknown distributions} \label{sim2}
We compare the performances of the two different approaches to the
spectrum access problem with unknown distributions that we discussed
in Section \ref{unknown}. We use a parameterized version of the
observation model we used in the example in Section \ref{sim1}. We
assume that the primary and secondary users are stationary and
assume that the secondary user is unaware of its location relative
to the primary transmitter. We assume that the secondary user
employs some form of energy detection, which means that the lack of
knowledge about the location of the primary manifests itself in the
form of an unknown mean power of the signal from the primary. Using
Gaussian distributions as in Section \ref{sim1}, we model the lack
of knowledge of the received primary power by assuming that the mean
of the observation under $H_1$ on channel $a$ is an unknown
parameter ${\theta_a}$ taking values in a finite set of positive
values $\Theta$. The new hypotheses are:
\begin{eqnarray}
S_a(k) = 0&:& {{Y_a(k)}} = {N_a(k)} \nonumber \\
S_a(k) = 1&:& {{Y_a(k)}} = {\theta_a} + {N_a(k)}  \nonumber \\ 
\mbox{ where } {N_a(k)} &\sim& \mathcal{N}({0}, \sigma^2), \theta_a
\in \Theta, \min(\Theta) > 0. \label{theta}
\end{eqnarray}

For the set of parameterized distributions in (\ref{theta}),
the log-likelihood ratio function $\mathcal{L}_{\theta}(x)$
defined in (\ref{llr}) is linear in $x$ for all $\theta \in
\Theta$. Hence comparing $\mathcal{L}_{\theta}(Y_a(k))$ to a
threshold is equivalent to comparing $Y_a(k)$ to some other
threshold. Furthermore, for this set of parameterized
distributions, it is easy to see that the conditional
cumulative distribution function (cdf) of the observations
$Y_a(k)$ under $H_1$, conditioned on $\theta_a$ taking value
$\theta$, is monotonically decreasing in $\theta$. Furthermore,
under the assumption that $\min \Theta > 0 $, it follows that
choosing $\theta^* = \min{\Theta}$ satisfies the conditions of
(\ref{condition}). Hence the optimal access decision under the
worst-case approach given in (\ref{accessworstnew}) can be
written as
\begin{equation}
\hat \delta_a(k) = \mathcal{I}_{\{u_k = a \}} \mathcal{I}_{\{Y_a(k)
< \tau_w \} }\label{accessworstexample}
\end{equation}
where $\tau_w$ satisfies
\begin{equation}
\mathsf{P}(\{Y_a(k) < \tau_w\} | \{S_a(k) = 1, \theta_a = \theta^*
\}) = \zeta 
\end{equation}
where $\theta^* = \min{\Theta}$. Thus the worst-case solution
for this set of parameterized distributions is identical to
that obtained for the problem with known distributions
described in (\ref{hypotheses}) with $\mu$ replaced by
$\theta^*$. Thus the structures of the access policy, the
channel selection policy, and the belief update equations are
identical to those derived in the example shown in Section
\ref{sim1} with $\mu$ replaced by $\theta^*$.

Similarly, the access policy for the case of random $\theta_a$
parameters given in (\ref{accessrandom}) can now be written as
\begin{equation}
\tilde \delta_a(k) = \mathcal{I}_{\{u_k = a \}}
\mathcal{I}_{\{Y_a(k) < \tau_r(k)\}  } \label{accessrandomexample}
\end{equation}
where $\tau_r(k)$ satisfies
\begin{equation}
\mathsf{P}(\{Y_a(k) < \tau_r(k)\} | \{S_a(k) = 1, \theta_a =
\theta^\#(k) \}) = \zeta 
\end{equation}
where $\theta^\#(k) = \min \Theta_a(k)$. The belief updates and
greedy channel selection are performed as described in Section
\ref{modeltheta}. The quantity $\epsilon_a(k)$ appearing in
(\ref{selectionrandom}) can now be written as
\[\epsilon_a(k) = \mathsf{P}(\{Y_a(k)
> \tau_r(k)\}| \{S_a(k) = 0\}).\]

We simulated the performances of both the schemes on the
hypotheses described in (\ref{theta}). We used the same values
of $L$, $P$, $\alpha$ and $\sigma$ as in Section \ref{sim1}. We
chose set $\Theta$ such that the SNR values in dB given by $20
\log \frac{ \mu_i}{\sigma}$ belong to the set $\{-5, -3, -1, 1,
3, 5\}$. The prior probability distribution for $\theta_a$ was
chosen to be the uniform distribution on $\Theta$. The
interference constraint $\zeta$ was set to $0.01$. Both
channels were assumed to have the same values of true SNR while
the simulations were performed. The reward was computed over
$10000$ slots since the remaining slots do not contribute
significantly to the reward. The value of $C$ in
(\ref{selectionrandommod}) was set to a value higher than the
number of slots considered so that the greedy channel selection
policy always uses the second alternative in
(\ref{selectionrandommod}). Although we require
(\ref{selectionrandommod}) for our proof of convergence of the
a posteriori probabilities in the Appendix, it was observed in
simulations that this condition was not necessary for
convergence.

The results of the simulations are given in \figref{figsim2}.
The net reward values obtained under the worst-case design of
Section \ref{worstcase} and that obtained with the algorithm
for learning $\theta_a$ given in Section \ref{modeltheta} are
plotted. We have also included the rewards obtained with the
greedy algorithm $G_1$ with known $\theta_a$ values; these
values denote the best rewards that can be obtained with the
greedy policy when the parameters $\theta_a$ are known exactly.
Clearly, we see that the worst-case design gives us almost no
improvement in performance for high values of actual SNR. This
is because the threshold we choose is too conservative for high
SNR scenarios leading to several missed opportunities for
transmitting. The minimal improvement in performance at high
SNR is due to the fact that the system now has more accurate
estimates of the channel beliefs although the update equations
were designed for a lower SNR level. The learning scheme, on
the other hand, yields a significant performance advantage over
the worst-case scheme for high SNR values as seen in the
figure. It is also seen that there is a significant gap between
the performance with learning and that with known $\theta_a$
values at high SNR values. This gap is due to the fact that the
posteriori probabilities about the $\theta_a$ parameters take
some time to converge. As a result of this delay in convergence
a conservative access threshold has to be used in the initial
slots thus leading to a drop in the discounted infinite horizon
reward. However, if we were using an average reward formulation
for the dynamic program rather than a discounted reward
formulation, we would expect the two curves to overlap since
the loss in the initial slots is insignificant while computing
the long-term average reward.

\begin{figure}
\centering
\includegraphics[width=7in]{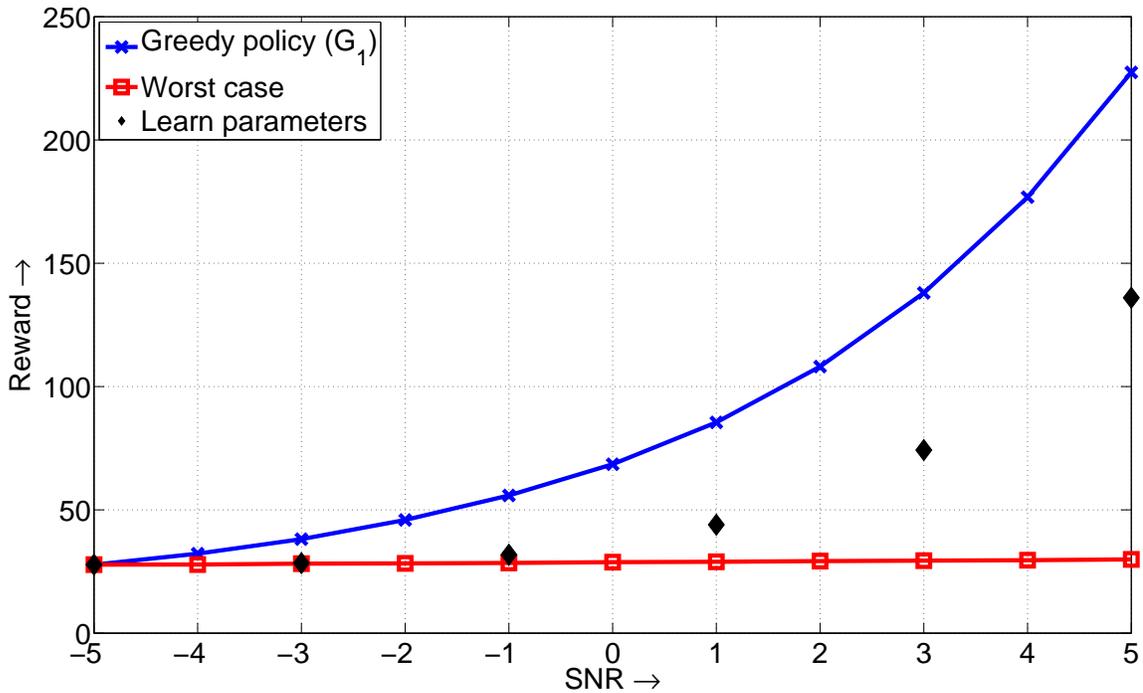}
\caption{Comparison of performances obtained with worst-case
approach and learning approach. Performance obtained with greedy
policy is also shown. The value of $\zeta =
0.01$.} \figlbl{figsim2}
\end{figure}

\begin{rem}
So far in this paper, we have assumed that the cardinality of
set $\Theta$ is finite. The proposed learning algorithm can
also be adapted for the case when $\Theta$ is a compact set. A
simple example illustrates how this may be done. Assuming
parameterized distributions of the form described in
(\ref{theta}), suppose that the value of $\theta_a$ in dB is
uniformly distributed in the interval $[-4.5, 4.5]$ and that we
compute the posteriori probabilities of $\theta_a$ assuming
that its value in dB is quantized to the finite set $\Theta =
\{-5, -4, \ldots, 5 \}$. Now if the actual realization of
$\theta_a$ is between $1 $ dB and $2 $ dB, say $1.5$ dB, then
we expect to see low posteriori probabilities for all elements
of $\Theta$ except $1 $ dB and $2 $ dB and in this case it
would be safe to set the access threshold assuming an SNR of $1
$ dB. Although this threshold is not the best that can be set
for the actual realization of $\theta_a$, it is still a
significant improvement over the worst-case threshold which
would correspond to an SNR of $-4.5$ dB. We expect the a
posteriori probabilities of all elements of $\Theta$ other than
$1$ dB and $2$ dB to converge to $0$, but the a posteriori
probabilities of these two values may not converge; they may
oscillate between $0$ and $1$ such that their sum converges to
$1$. A rigorous version of the above argument would require
some ordering of the parameterized distributions as in
(\ref{condition}).
\end{rem}

\section{Conclusions and discussion} \label{comments}
The results of Section \ref{sim1} and the arguments we
presented in Section \ref{comparison} clearly show that our
scheme of estimating the channel occupancies using the
observations yields performance gains and may have practical
advantages over the ACK-based scheme that was proposed in
\cite{chen_zhao_swami}. We believe that these advantages are
significant enough to justify using our scheme even though it
necessitates the use of dedicated control channels for
synchronization.

For the scenario where the distributions of the received
signals from the primary transmitters are unknown and belong to
a parameterized family, the simulation results in Section
\ref{sim2} suggest that designing for worst-case values of the
unknown parameters can lead to a significant drop in
performance relative to the scenario where the distributions
are known. Our proposed learning-based scheme overcomes this
performance drop by learning the primary signal's statistics.
The caveat is that the learning procedure requires a reliable
model for the state transition process if we need to give
probabilistic guarantees of the form (\ref{thetaguarantee}) and
to ensure convergence of the beliefs about the $\theta_a$
parameters.

In most of the existing literature on sensing and access policies
for cognitive radios that use energy detectors, the typical practice
is to consider a worst-case mean power under the primary-present
hypothesis. The reasoning behind this approach is that the cognitive
users have to guarantee that the probability of interfering with any
primary receiver located within a protected region \cite{somefund,
jstsp} around the primary transmitter is below the interference
constraint. Hence it is natural to assume that the mean power of the
primary signal is the worst-case one, i.e., the mean power that one
would expect at the edge of the protected region. However, the
problem with this approach is that by designing for the worst-case
distribution, the secondary users are forced to set conservative
thresholds while making access decisions. Hence even when the
secondary users are close to the primary transmitter and the SNR of
the signal they receive from the primary transmitter is high, they
cannot efficiently detect vacancies in the primary spectrum.
Instead, if they were aware that they were close to the transmitter
they could have detected spectral vacancies more efficiently as
demonstrated by the improvement in performance at higher SNRs
observed in the simulation example in Section \ref{sim1}. This loss
in performance is overcome by the learning scheme proposed in
Section \ref{modeltheta}. By learning the value of $\theta_a$ the
secondary users can now set more liberal thresholds and hence
exploit vacancies in the primary spectrum better when they are
located close to the primary transmitter. Thus, using such a scheme
would produce a significant performance improvement in overall
throughput of the cognitive radio system.

\appendix
Here we show that for each channel $a$, the a posteriori
probability mass function of parameter $\theta_a$ converges to
a delta-function at the true value of the parameter almost
surely under the algorithm described in Section
\ref{modeltheta}.

\begin{thm}
Assume that the transition probability matrix $P$ satisfies
(\ref{assnew}). Further assume that the conditional densities
of the observations given in (\ref{parameter}) satisfy
\begin{equation}
\int f_{\mu_i}(y) |\log (f_{\mu_i} (y))| dy < \infty \mbox{ for all }\mu_i \in \Theta, \label{mutualabscont}
\end{equation}
and that all densities in (\ref{parameter}) are distinct. Then,
under the channel sensing scheme that was introduced in
(\ref{selectionrandommod}), for each channel $a$,
\[
\mathsf{P}(\{\theta_a = \mu_i\}|Y^n) \xrightarrow[n \to
\infty]{a.s.} \mathcal{I}_{\{\theta_a =\mu_i\}}, \mbox{
for all } \mu_i \in \Theta.
\]
\end{thm}
\begin{proof}
Without loss of generality, we can restrict ourselves to the
proof of the convergence of the a posteriori distribution of
$\theta_1$, the parameter for the first channel. By the
modified sensing scheme introduced in
(\ref{selectionrandommod}), it can be seen that channel $1$ is
sensed at least every $ML$ slots. Hence, if the a posteriori
distribution converges for an algorithm that senses channel $1$
exactly every $ML$ slots, it should converge even for our
algorithm, since our algorithm updates the a posteriori
probabilities more frequently. Furthermore by considering an
$ML$-times undersampled version of the Markov chain that
determines the evolution of channel $1$, without loss of
generality, it is sufficient to show convergence for a sensing
policy in which channel $1$ is sensed in every slot. It is
obvious that since condition (\ref{assnew}) holds for the
original Markov chain, it holds even for the undersampled
version. So now we assume that an observation $Y_k$ is made on
channel $1$ in every slot $k$. We use $Y^k$ to represent all
observations on channel $1$ up to slot $k$.

We use $\mu_{i^*} \in \Theta$ to represent the true realization
of random variable $\theta_1$ with $i^* \in \{1,\ldots,N\}$,
and $\pi$ to denote the prior distribution of $\theta_1$. The a
posteriori probability mass function of $\theta_1$ evaluated at
$\mu_i$ after $n$ time slots can be expressed as
\begin{eqnarray}
\mathsf{P}(\{\theta_1 = \mu_j\}|Y^n) &=& \frac{P_{j}(Y^n) \pi(\mu_{j})}{ \sum_i P_i(Y^n) \pi(\mu_{i})}
\label{posteriori0}
\end{eqnarray}
where we use the notation $P_i(.)$ to denote the distribution
of the observations conditioned on $\theta_1$ taking the value
$\mu_i \in \Theta$. It follows from \cite[Theorem 1, Theorem 2,
and Lemma 6]{leroux92} that conditioned on $\{\theta_1 =
\mu_{i^*}\}$ we have,
\[
\frac{P_{i^*}(Y^n)}{P_i(Y^n)} \xrightarrow[n \to \infty]{a.s.} \infty \mbox{ for all } i \neq i^*.
\]
Hence, it follows from (\ref{posteriori0}) that conditioned on
$\{\theta_1 = \mu_{i^*}\}$ we have,
\[
\frac{P_{i^*}(Y^n) \pi(\mu_{i^*})}{ \sum_i P_i(Y^n)  \pi(\mu_{i})} \xrightarrow[n \to \infty]{a.s.} 1
\]
which further implies that conditioned on $\{\theta_1 =
\mu_{i^*}\}$ we have,
\[
\frac{P_{j}(Y^n) \pi(\mu_{j})}{ \sum_i P_i(Y^n)\pi(\mu_{i})} \xrightarrow[n \to \infty]{a.s.} \mathcal{I}_{\{i^* = j\}}.
\]
Since this holds for all possible realizations $\mu_{i^*} \in
\Theta$ of $\theta_1$, the result follows.
\end{proof}

\section*{Acknowledgment}
The authors would like to thank Prof. Vivek Borkar for
assistance with the proof of the convergence of the learning
algorithm in Section \ref{modeltheta}.

\end{document}